\documentclass[aps,prd,onecolumn, showpacs,floatfix]{revtex4}
\usepackage{amssymb,amsmath,epsfig,graphicx}
\usepackage{caption}
\usepackage{dcolumn}

\newcommand{\be}{\begin{equation}}
\newcommand{\ee}{\end{equation}}
\newcommand{\bea}{\begin{eqnarray}}
\newcommand{\eea}{\end{eqnarray}}
\newcommand{\bi}{\begin{itemize}}
\newcommand{\ei}{\end{itemize}}

\begin{document}

\title{\bf Quantum slow-roll and  quantum fast-roll
  inflationary initial conditions: CMB quadrupole suppression and
further effects on the low CMB multipoles}
\author{\bf F. J. Cao$^{(a,c)}$} \email{francao@fis.ucm.es}
\author{\bf H. J. de Vega$^{(b,c)}$} \email{devega@lpthe.jussieu.fr}
\author{\bf N. G. Sanchez$^{(c)}$} \email{Norma.Sanchez@obspm.fr}
\affiliation{ $^{(a)}$ Departamento de F\'{\i}sica At\'omica, Molecular y
Nuclear, \\ Universidad Complutense de Madrid, \\ Avenida Complutense s/n,
28040 Madrid, Spain. \\
$^{(b)}$ LPTHE, Laboratoire Associ\'e au CNRS UMR 7589,\\
Universit\'e Pierre et Marie Curie (Paris VI) et Denis Diderot (Paris VII),\\
Tour 24, 5 \`eme. \'etage, 4, Place Jussieu, 75252 Paris, Cedex 05, France.\\
$^{(c)}$ Observatoire de Paris, LERMA, Laboratoire Associ\'e au CNRS UMR 8112,
 \\61, Avenue de l'Observatoire, 75014 Paris, France.}

\date{\today}

\begin{abstract}
Quantum fast-roll initial conditions for the inflaton which are
different from the classical fast-roll conditions and from the
quantum slow-roll conditions can lead to inflation that last long
enough. These quantum fast-roll initial conditions for the
inflaton allow for kinetic energies of the same order of the
potential energies and nonperturbative inflaton modes with nonzero
wavenumbers. Their evolution starts with a transitory epoch where
the redshift due to the expansion succeeds to assemble the quantum
excited modes of the inflaton in a homogeneous (zero mode)
condensate, and the large value of the Hubble parameter succeeds
to overdamp the fast-roll of the redshifted inflaton modes. After
this transitory stage the effective classical slow-roll epoch is
reached. Most of the efolds are produced during the slow-roll
epoch and we recover the classical slow-roll results for the
scalar and tensor metric perturbations plus corrections. These
corrections are important if scales which are horizon-size today
exited the horizon by the end of the transitory stage and as a
consequence the lower CMB multipoles get suppressed or enhanced.
Both for scalar and tensor metric perturbations, fast-roll leads
to a {\bf suppression} of the amplitude of the perturbations (and
of the low CMB multipoles), while the quantum precondensate epoch
gives an {\bf enhancement} of the amplitude of the perturbations
(and of the low CMB multipoles). These two types of corrections
can compete and combine in a scale dependent manner. They turn to
be smaller in new inflation than in chaotic inflation. These
corrections arise as natural consequences of the quantum
nonperturbative inflaton dynamics, and can allow a further
improvement of the fitting of inflation plus the $\Lambda$CMB
model to the observed CMB spectra. In addition, the corrections to
the tensor metric perturbations will provide an independent test
of this model. Thus, the effects of quantum inflaton fast-roll
initial conditions provide a consistent and contrastable model for
the origin of the suppression of the quadrupole and for other
departures of the low CMB multipoles from the slow-roll
inflation-$\Lambda$CMB model which are to be contrasted to the TE
and EE multipoles and to the forthcoming and future CMB data.
\end{abstract}

\maketitle

\tableofcontents

\section{Introduction} \label{sec:intro}

Inflation  (an epoch of accelerated expansion of the universe)
solves the horizon and flatness problems of the
standard Big Bang model. It naturally generates scalar density
fluctuations that seed large scale
structure and the temperature anisotropies in the cosmic
microwave background (CMB),
and tensor perturbations (primordial gravitational waves)
\cite{guth,infbooks}. Inflation is
based on a scalar field (the inflaton) whose homogeneous expectation
value drives the dynamics of the
scale factor, plus small quantum fluctuations.

\medskip

A great diversity of inflationary models predict fairly generic
features: a gaussian, nearly scale invariant spectrum of (mostly)
adiabatic scalar and tensor primordial fluctuations,  these
provide an excellent fit to the highly precise wealth of data
of the Wilkinson Microwave Anisotropy Probe (WMAP)
\cite{wmap} making the inflationary paradigm fairly robust.
Precise CMB data reveal peaks and valleys in the temperature
fluctuations resulting from  acoustic oscillations in the
electron-photon fluid at recombination. These and future CMB and
large scale structure (LSS) observations require more precise
theoretical predictions from inflation, and a deeper understanding
of how inflation begins and ends.

\medskip

Amongst the wide variety of inflationary scenarios,
single field slow-roll
models provide an appealing, simple and fairly generic description of
inflation. Its simplest implementation is based on a single
scalar field (the inflaton). The inflaton potential is
fairly flat and it dominates the universe energy during
inflation. This flatness leads to a slowly
varying Hubble parameter (slow-roll) ensuring a sufficient number
of inflation e-folds
to explain the homogeneity, isotropy and flatness of the universe,
and also explains the gaussianity of the fluctuations
as well as the (almost) scale invariance of their power spectrum.

\medskip

The dynamics of inflation is usually described by the classical
evolution of the inflaton in a dynamical space-time whose scale
factor obeys the classical Friedmann equation. This is justified
by the enormous stretching of physical lengths during inflation
that classicalizes the dynamics. On the other hand, both the
perturbations of the metric and of the inflaton are computed
considering their quantum nature. The observable consequences of
slow-roll inflation are well known \cite{infbooks}, in particular
those for the CMB anisotropies. Present observational data agree
with its predictions for a class of slow-roll
inflaton potentials except for the CMB quadrupole suppresion \cite{wmap,cobe}.

However, the large energy densities during inflation allows for
the creation of inflaton particles, and call for a quantum
treatment for the inflaton field. Refs.~\cite{tsuinf,cao04,qnew}
considered a quantum treatment to study the dynamics of the
inflaton. Quantum excited initial conditions lead to inflation
that last long enough provided the quantum inflaton background
satisfies the quantum generalized slow-roll conditions
\cite{tsuinf,cao04}. In addition, after a transitory precondensate
epoch, the excited modes of the inflaton assembles in an
homogeneous zero mode condensate \cite{qnew,tsuinf,cao04}. Once
formed, this condensate obeys the inflaton classical evolution
equations and we therefore recover the classical inflation
slow-roll description.

\medskip

More recently, classical fast-roll initial conditions have been shown to lead to
long enough inflation and to the CMB quadrupole suppresion
\cite{boy06,des08,lasenby}. In this article we consider
more general inflaton initial conditions which lead to long enough
inflation, and that we call quantum fast-roll initial conditions.
They are different from the
quantum slow-roll conditions and include the classical fast-roll
conditions as a particular case.

All these general initial conditions lead after a transitory
precondensate stage
to a slow roll inflationary epoch that admits a classical description thus
recovering the classical slow-roll inflation predictions for the scalar
and tensor metric perturbations. Moreover, if the larger cosmologically
relevant scales exited the horizon close to the end of the transitory
stage, the effect of the modified initial conditions in the bulk modes
that dominate the energy density will be imprinted
as corrections to the amplitudes of the lower CMB multipoles.
These corrections have been computed for classical fast
roll initial conditions \cite{boy06}, and they have been shown to
improve the fit of inflation plus the $\Lambda$CMB
model to the CMB data \cite{des08}. For
quantum slow-roll initial conditions the corrections were first estimated
in ref.~\cite{cao04} and they will be computed in more detail here.

In the present article we also compute the corrections due to the quantum
fast-roll initial conditions introduced here.
We show that quantum slow-roll conditions and quantum fast-roll conditions
may improve the fit of the CMB data with the $\Lambda$CMB model plus
inflation.These corrections to the spectrum of scalar and tensor metric
perturbations arise as natural consequences of the quantum precondensate
dynamics of the inflaton field. The two
possible types of transitory epochs present before slow-roll, namely,
effective classical fast-roll and quantum
(fast-roll or slow-roll) precondensate, lead to opposite kinds of
corrections for the low CMB multipoles.
This quantum  precondensate dynamics predict
corrections to the low TE and EE multipoles (temperature T and
E-polarization mode correlations) and to the low CMB tensor multipoles.
Therefore, this is a consistent and contrastable theory for the
origin of the suppression of the quadrupole and for other departures of
the low CMB multipoles data from the classical slow-roll inflation plus
the $\Lambda$CMB model.

\medskip

This paper is organized as follows:
in section \ref{sec:qie} we present the quantum inflaton field and its
evolution equations in the limit of large number of components, as well
as the theory to compute corrections to the  spectrum of
perturbations due to departures from classical
slow-roll evolution, and the relative change in the CMB quadrupole due to
these corrections. In Section \ref{sec:qic}, we compute first the corrections to the
primordial spectrum of perturbations due to quantum slow-roll initial conditions.
Next, we introduce quantum fast roll initial conditions and compute the
inflationary evolution. Then, we compute the  corrections to the
primordial spectrum of perturbations due to the quantum fast
roll initial conditions, and the relative change in the low CMB multipoles.
Finally, in section IV we further discuss the results and present our conclusions.

We use units $ \hbar = 1 $, $ c =1 $, and in the figures we also
take $ m = 1 $ with $ m $ the inflaton mass. For the scale factor of the FRW
metric we take $ a(0) = 1 $.

\section{Quantum inflaton evolution} \label{sec:qie}

The action for the {\em quantum inflaton dynamics} can be written as,
\be
  S = S_g + S_m + \delta S_g + \delta S_m
\ee
where each term describes the dynamics of one of the components:
$ S_g $ describes the dynamics of the metric background, $ \delta S_g $ that
of the cosmologically relevant metric perturbations, $ S_m $ accounts
for the dynamics
of the inflaton field, and $ \delta S_m $ the dynamics of the
cosmologically relevant inflaton perturbations. We consider
here all these terms quantum mechanical except for the metric background
$ S_g $ which is purely classical. This generalizes the classical framework
where only the perturbations are quantized.

On one hand, the gravitational terms are
\be \label{gravact}
  S_g + \delta S_g = -\frac{1}{16\pi G} \int{\sqrt{-g}\;d^4x\,R}
\ee
where $ G $ is Newton's gravitational constant, and $ R $
is the Ricci scalar for the complete metric $ g_{\mu\nu} $.
By expanding $ g_{\mu\nu} $ in terms of the spatially flat FRW metric and
its perturbation, $ g_{\mu\nu} = g_{\mu\nu}^{(FRW)} + \delta
g_{\mu\nu} $, the corresponding $ S_g $ terms (those which do not
contain $ \delta g_{\mu\nu} $) and the $ \delta S_g $ terms
can be recognized (see refs.~\cite{cao04,revmuk} for more details). We are
interested in the metric perturbations at the cosmologically relevant
scales.

On the other hand, the inflaton terms in the action are
\be \label{qftact}
  S_m + \delta S_{m} = \int{\sqrt{-g}\;d^4x
    \left[\frac12\,\partial_{\alpha}\vec{\chi}\,
    \partial^{\alpha}\vec{\chi} - V(\vec{\chi}) \right]}
\ee
where $ \vec \chi $ is the inflaton quantum field and $
V(\vec{\chi}) $ the inflaton potential. We consider a
multicomponent inflaton quantum field $ \vec \chi = (\chi_1,
\;\ldots, \;\chi_N) $ since it allows to apply a nonperturbative
method to compute the dynamics, namely the large $N$ limit. It
must be stressed that we consider $ O(N) $ invariant inflaton
potentials, and $ O(N) $ invariant initial states as the universe
as a whole should be expected to be in a $ O(N) $ invariant state.
This leads to straight trajectories in field space and therefore
this is effectively a single field inflaton model \cite{cao04}.

It is also convenient to split the contribution from the inflaton terms
in the action $ S_m $ and the contributions from the cosmologically
relevant inflaton and metric perturbations in the term $ \delta S_m $.

\medskip

Let us call $ \Lambda $ the $k$-scale in momentum space that
separates the modes contributing to the background ($ k < \Lambda
$) from those that contribute to the perturbations ($ k > \Lambda
$). Modes with $ k \gg m $ cannot be significantly excited since
the energy density during inflation must be of the order $ \sim 20
\; M^4 $ in order to have $ \gtrsim 60 $ efolds of inflation
\cite{tsuinf} where $ M \sim 10^{16}$ GeV is the energy scale of
inflation,
\be \label{masinf}
m = \frac{M^2}{M_{Pl}}
\ee
is the inflaton mass, and $ M_{Pl} $ is the Planck mass given by $
M_{Pl} = 1/\sqrt{8\pi G} = 2.4 \times 10^{18}$ GeV.

On the other hand, the modes that are cosmologically relevant,
i.e., those corresponding to the scales of large scale
structures and the CMB are today in the range from $ 0.4 $ Mpc
to $ 4 \times 10^3 $ Mpc . These scales correspond to physical
wavenumbers at the beginning of inflation in the range from \cite{des08}
\be \label{rango}
10 \; e^{N_Q-62} \; m  \; \sqrt{\frac{H}{10^{-4} \; M_{Pl}}}
\quad {\rm to}  \quad 10^5 \; e^{N_Q-62} \; m \;
\sqrt{\frac{H}{10^{-4} \; M_{Pl}}} \; ,
\ee
where $ N_Q $ is the number of efolds since the CMB quadrupole modes exit
the horizon till the end of inflation
and $ H $ is the Hubble parameter during inflation.

Notice that the CMB quadrupole modes are today horizon size
($ 4 \times 10^3 $ Mpc) while at horizon exit their physical wavenumber
corresponds to the left hand side of eq.(\ref{rango}).

Thus, $ \Lambda $ is in an intermediate $k$-range of modes whose
contributions are
irrelevant both for the background and for the perturbations at the
cosmologically relevant scales, $ \Lambda \sim 10 \; e^{N_Q-62} \; m  \;
\sqrt{\frac{H}{10^{-4} \; M_{Pl}}} $ and the observable results
are independent
of the particular value of $ \Lambda $. (For a more detailed discussion see
ref.~\cite{cao04}).

\medskip

The quantum evolution equation for the inflaton background is
\be \label{qeib}
\ddot{\vec{\chi}} + 3 H \dot{\vec{\chi}} - \frac{\nabla^2\vec{\chi}}{a^2}
  + V'(\vec{\chi}) = 0
\ee
with $ a $ the scale factor of the FRW metric, $ H \equiv \dot a / a $ the
Hubble parameter, and $ V' $ the derivative of the inflaton potential
with respect to $ \vec \chi $ and the dot $ \dot{} $ stands for the
derivative with respect the cosmic time.
As the background is homogeneous, this implies that the expectation values
of the field and of its momenta are homogeneous.
It is important to note that this homogeneity of the
expectation values does not prevent to have excited modes with nonzero
wavenumbers. These homogeneous states with wavenumbers excited modes
can be thought as homogeneous seas of particles with nonzero momenta.

The evolution equations for the quantum inflaton are coupled to
the Friedmann classical evolution equations for the space-time metric
\be
G_{\mu\nu} = \langle T_{\mu\nu} \rangle \implies H^2 = \frac{\rho}{3M_{Pl}^2}
\ee
with $ G_{\mu\nu} $ the Einstein tensor for the FRW metric, $
\langle T_{\mu\nu} \rangle $ the expectation value of the
energy-momentum tensor for the inflaton quantum field, and $ \rho
= \langle T^{00} \rangle $ is the energy density.

\medskip

The approach of considering the expectation value of the
energy-momentum tensor of the matter quantum fields as the source
for the metric evolution is usually called semiclassical gravity.
This approach is justified provided the quantum gravity
corrections can be neglected, as is our case here: they are of
order $ \sim \left(m / M_{Pl}\right)^2 \sim 10^{-9} $.

\medskip

The evolution equations for the quantum perturbations of the
metric and of the inflaton are well known when the inflaton
is a classical field (see for example ref.~\cite{revmuk}). However, they are
not known in the case of quantum inflaton fields. It is only
known how to compute the spectrum of perturbations in
the regimes where the quantum inflaton can be described by an effective
classical field.

\subsection{Quantum evolution equations. The precondensate and the condensate epochs.}

It is not an easy task to compute the time evolution defined by
eq.~\eqref{qeib} for a generic
quantum state and it is only know how to compute this
evolution under certain approximations. Here, as we need
to compute the evolution of excited nonperturbative states with homogeneous
expectation values, we will use the large $ N $ limit method
\cite{tsuinf,cao04,qnew,tsu}. This method
gives the evolution equations in the limit of large number of components
$ N $ and provides consistent evolution equations which can be numerically
implemented both for the inflaton expectation value and for the
nonperturbative inflaton modes. The large $N$ limit captures
the relevant physics of inflation.

We consider the inflaton potential
\be
  V(\vec{\chi}) = \frac12\; \alpha \; m^2\,\vec{\chi}^2
    + \frac{\lambda}{8N}\,\left( \vec{\chi}^2 \right)^2 +
    \frac{Nm^4}{2\lambda}\frac{1-\alpha}{2}\;, \label{potential}
    \quad \quad \mbox{ with } \alpha = \pm 1
    \;,
\ee
The last constant term here sets the minimum of $ V(\vec{\chi}) $ at
$ V(\vec{\chi}) = 0 $. The quartic selfcoupling $ \lambda $ is of the
order \cite{1sN}
\be\label{lam}
\lambda \sim \left(\frac{M}{M_{Pl}}\right)^4 \sim 10^{-12}
\ee
The quantum field Fock state (or density matrix) describing the inflaton field
is assumed invariant under traslations and rotations. That is, homogeneous and
isotropic \cite{tsuinf,qnew}. The quantum field states consist on a distribution
of $k$-modes over the Fock vacuum. We choose a mode distribution of width
$ \Delta k $ centered around a value $ k = k_0 < \Lambda $.
The field operator $ \vec{\chi}(\vec x, t) $ can be Fourier expanded in terms
of creation and annihilation operators and mode functions
\cite{cao04}-\cite{tsuinf} as
\begin{equation}\label{oper}
\vec{\chi}(\vec x, t) = {\hat e}_1 \; \varphi(t) +
\int\frac{d^3k}{(2\pi)^3}
\left[{\vec a}_k \; f_k(t) \; e^{i\vec{k}\cdot \vec x} + {\vec a}^{\dagger}_k
 \; f^*_k(t) \; e^{-i\vec{k}\cdot \vec x} \right] .
\end{equation}
where $ {\hat e}_1 =(1,0,\ldots,0) $ and
$ {\vec a}_k $ and $ {\vec a}^{\dagger}_k $ obey canonical commutation rules.

In the large $ N $ limit the evolution equations for
the quantum inflaton background are given by \cite{cao04,tsuinf}
\begin{eqnarray}
&&\ddot\varphi + 3\,H\,\dot\varphi + {\cal M}^2\,\varphi = 0
  \label{eqNexpect} \\
&&\ddot f_k +3\,H\,\dot f_k + \left(\frac{k^2}{a^2} + {\cal
    M}^2\right)\,f_k = 0  \label{eqNmodes} \\
&&\mbox{with }\quad {\cal M}^2 = \alpha \; m^2 + \frac{\lambda}{2}\,\varphi^2 +
\frac{\lambda}{2}\,\int_R{\frac{d^3k}{2(2\pi)^3}\,|f_k|^2}  \; ,
\label{masaef}
\end{eqnarray}
where $ \varphi \equiv <\vec{\chi}_1> $ is the expectation value of the
quantum inflaton field, $ <\vec{\chi}_a> = 0 $ for $ 2 \leq a \leq N $
and $ f_k $ its quantum modes \cite{cao04}.
The inflaton evolution equations
are coupled to the evolution equations for the scale factor,
\be
H^2 = \frac{\rho}{3\,M_{Pl}^2} \quad \; ,
\quad \quad \frac{\rho}{N} = \frac12 \, \dot\varphi^2 + \frac{{\cal
M}^4 - m^4}{2\lambda} + \frac{m^4}{2\lambda}\frac{1-\alpha}{2} +
\frac14 \int_R \frac{d^3k}{(2\pi)^3} \left(|\dot f_k|^2 +
\frac{k^2}{a^2}|f_k|^2\right)\;. \label{H2rho}
\ee
where $ \rho = \langle T^{00} \rangle $ is the energy density. The
pressure ($ p \; \delta_i^{\;j} = \langle T_i^{\;j} \rangle $) is
given by
\be
\frac{p+\rho}{N} = \dot\varphi^2 + \frac12 \int_R
\frac{d^3k}{(2\pi)^3} \left(|\dot f_k|^2 +
\frac{k^2}{3a^2}|f_k|^2\right)\; . \label{pressure}
\ee
The index $ R $ denotes the renormalized expressions of these
integrals \cite{tsuinf}. This means that we must subtract the appropriate
asymptotic ultraviolet behavior in order to make convergent
the integrals in eqs.(\ref{masaef})-(\ref{pressure}).

The characteristic mass scale of the inflaton modes $ f_k(t) $ is clearly
the inflaton mass $ m $ [see eqs.(\ref{eqNmodes})-(\ref{masaef})].
In order to have an energy density $ \rho = {\cal O}(M^4) $
and not $ {\cal O}(m^4) \ll {\cal O}(M^4) $, the inflaton modes
with $ k < \Lambda $ must have a non-perturbative amplitude
$ |f_k(t)| = {\cal O}(\lambda^{-1/2}) $. Thus,
$$
\rho = {\cal O}\left(\frac{m^4}{\lambda}\right) =  {\cal O}(M^4) \; .
$$
where we used eqs.(\ref{masinf}), (\ref{lam}) and (\ref{H2rho}).

Two types or classes of quantum inflation scenarios have been
considered \cite{qnew,tsuinf,cao04}:
\begin{itemize}
\item{(i) Quantum chaotic
inflation, in which the inflaton field has a large amplitude (in
the expectation value and/or in the quantum modes). The
inflaton potential energy is due to the potential energy of this
large amplitude state.}
\item{(ii) Quantum new
inflation, in  which the inflaton field has a small amplitude, and the
inflaton potential energy is due
to the large values of the potential for small amplitudes of the field.}
\end{itemize}

In this framework the quantum generalized slow-roll condition is given by
\cite{tsuinf,cao04}
\be \label{gsrc}
\dot\varphi^2+\int_R{\frac{d^3k}{2(2\pi)^3}\,|\dot f_k|^2} \;\ll\; m^2
\left( \varphi^2 + \int_R{\frac{d^3k}{2(2\pi)^3}\,|f_k|^2} \right)
\ee
This is a sufficient condition to guarantee inflation
($ \ddot a > 0 $) and that it lasts long (for both scenarios). The classical
slow-roll condition is a particular case
\be \label{csrc}
 \dot\varphi^2 \ll m^2 \varphi^2.
\ee
The states that verify the quantum generalized slow-roll conditions
eq.\eqref{gsrc} lead to two inflationary epochs separated
by a condensate formation \cite{tsuinf,cao04},

\begin{itemize}
\item (a) {\bf The pre-condensate epoch}: During this epoch the term
\be \label{mome}
\int_R \frac{d^3k}{(2\pi)^3} \; \frac{k^2}{a^2} \; |f_k|^2
\ee
gives an important contribution to the energy [eq.~\eqref{H2rho}] while it
decreases due to the redshift of the excitations ($ k/a \to 0 $).
This epoch ends when
all the contributions of $ k^2/a^2 $ to the background dynamics are
negligible. Then, all the excited modes are effectively assembled in
a zero mode condensate.

\item (b) {\bf The condensate slow-roll epoch}: Once the excited modes
behave effectively as a zero mode condensate, the inflaton
background can be described as the classical effective field
\cite{qnew,cao04}
\be \label{efffield}
\tilde{\varphi}_{eff}(t) = \sqrt{N}
\sqrt{\varphi^2(t)+\int{\frac{d^3k}{2(2\pi)^3}\,|f_k(t)|^2}}\;.
\ee
This effective classical inflaton verifies the classical evolution equations
\bea
&&\ddot{\tilde{\varphi}}_{eff} + 3 \,H \,\dot{\tilde{\varphi}}_{eff}
  + \alpha \; \tilde m^2 \; \tilde{\varphi}_{eff} + \tilde{\lambda}\;
  \tilde{\varphi}_{eff}^3 = 0\;, \label{eqclasphi} \\
&&H^2 = \frac{\tilde \rho}{3 M_{Pl}^2}\;, \quad \quad
 \tilde \rho = \frac12\; \dot{\tilde{\varphi}}_{eff}^2
 + \frac12\; \alpha \; \tilde m^2 \; \tilde{\varphi}_{eff}^2 +
 \frac{\tilde{\lambda}}{4}\;\tilde{\varphi}_{eff}^4 +
\frac{m^4}{8 \; \tilde{\lambda}}
(1 - \alpha) \; , \label{epsilonclas} \\
&&\mbox{with } \quad  \tilde{\lambda} = \frac{\lambda}{2 N} \; ,
  \quad \tilde m^2 = m^2 \; , \quad \alpha = \pm 1 \; .
  \eea
The pressure is given by
\be \label{pclas}
\tilde p + \tilde \rho = \dot{\tilde{\varphi}}_{eff}^2 \;.
\ee
It is important to note that the initial conditions for the
classical effective inflaton $ \tilde{\varphi}_{eff} $
are fixed by the quantum state (i.e. the quantum precondensate epoch).
During the
condensate slow-roll epoch the effective inflaton $ \tilde{\varphi}_{eff} $
verifies the classical slow-roll condition eq.\eqref{csrc}, i.e.,
$ \dot{\tilde{\varphi}}_{eff}^2 \ll m^2 \; {\tilde{\varphi}}_{eff}^2 $.

\end{itemize}

In addition, we introduce and study in this article novel quantum
initial conditions for the inflaton which are not of quantum slow-roll
type, and that we call quantum fast-roll initial conditions.
These quantum fast-roll
initial conditions also lead to inflation that last long enough,
as we show in the next section. The effective description
eqs.~\eqref{efffield}-\eqref{pclas} also reveals to be valid for
quantum fast-roll initial conditions after the evolution has
redshifted the modes and assembled them in a zero mode condensate.

These quantum initial conditions concern the nonperturbative inflaton
modes that contribute significantly to the background, that is the
$ k < \Lambda $ inflaton modes.
We choose the usual Bunch-Davies initial conditions for the cosmologically
relevant perturbations $ k > \Lambda $. Physical effects of
non-Bunch-Davies initial conditions for the cosmologically relevant
modes were presented in ref. \cite{boy06}.

\subsection{Corrections to the Spectrum of perturbations.
The transfer function $ D(k) $. }

The spectrum of scalar and tensor metric perturbations for quantum
inflation (new and
chaotic) has been computed in ref.~\cite{cao04}. There, we found
that provided the cosmologically relevant scales exited the horizon
during the
condensate epoch, the classical inflation results are recovered at
leading order. In addition, this study had identified several sources
of quantum corrections and computed their order of magnitude:
\begin{itemize}
\item{(i) coupling of the cosmological relevant modes with
the quantum inflaton dynamics yields corrections of order $ \sim  (H/M_{Pl})^2
\sim 10^{-9} $ or smaller, this estimation
has been confirmed by the more detailed study
in ref.~\cite{boysuperhorizon};}
\item{ (ii) contributions from higher orders in $ 1/N $;}
 \item{ (iii) contributions from the precondensate epoch. These can be important
for large scales that exited the horizon before or during the formation
of the condensate.}
\end{itemize}
Here we compute these last quantum corrections in more detail and
for more general initial conditions. That is, we compute the
effects of the quantum precondensate epoch in the primordial
spectrum of perturbations and in the observed CMB multipoles.

In addition to the usual method to compute the primordial spectrum
of perturbations, we also use here the method developed in
refs.~\cite{boy06} to compute the changes in the primordial
spectrum due to generic initial conditions, in particular the
fast-roll conditions. Both of these methods have been obtained for
a classical inflaton field. Therefore, they can only be applied
in principle to regimes where the quantum inflaton can be described
by an effective classical field, as in the condensate epoch. In
order to obtain estimates of the effects of the precondensate
epoch we use these formalisms when the inflaton is close to the
formation of the condensate and thus can be described
by the effective inflaton eq.~\eqref{efffield}.

The modes of the curvature perturbations $ S_{\cal  R} $ fullfil
the equation \cite{boy06}
\be\label{ecmod}
\left[ \frac{d^2}{d\eta^2} + k^2 - W_{\cal R}(\eta) \right]
S_{\cal  R}(k;\eta) = 0 \;, \quad \quad W_{\cal R}
= \frac{1}{z} \frac{d^2z}{d\eta^2}
\ee
where $ \eta \equiv \int_0^t dt'/a(t') $ is the conformal time and
$ z(\eta) $ is given in terms of the derivative of the classical
inflaton which for our study here is the effective inflaton field
$ \tilde{\varphi}_{eff}(t) $ eq.(\ref{efffield}), i.e.,
\be
z(\eta) = a \; \frac{\dot{\tilde{\varphi}}_{eff}}{H}.
\ee
$ W_{\cal R}(\eta) $ is the potential felt by the curvature
perturbations during the whole evolution: fast-roll as well as slow-roll.

For the curvature perturbations the quantity that leads to the differences
with respect to slow-roll is the potential
\be\label{potVR}
{\cal V}_{\cal R} = W_{\cal R} - W^{sr}_{\cal R}
\ee
that is, the difference of the potential $ W_{\cal R} $ and its
value for slow-roll, $ W^{sr}_{\cal R} $, where to first order in slow-roll,
\be\label{potW}
W^{sr}_{\cal R} = - \frac{2+9 \, \epsilon_v-3 \, \eta_v}{\eta^2} \; ,
\ee
with $ \epsilon_v, \; \eta_v $ the first order slow-roll parameters
evaluated during slow-roll:
$$
\epsilon_v = \frac{\dot{\tilde{\varphi}}_{eff}^2}{2 \, M_{Pl}^2 \; H^2}
\quad {\rm and} \quad
\eta_v = M_{Pl}^2 \; \frac{V''(\tilde{\varphi}_{eff})}{
V(\tilde{\varphi}_{eff})} \; .
$$
The potential $ {\cal V}_{\cal R} $ eq.(\ref{potVR}) is determined
by the departures of the background dynamics from slow-roll.
Notice that when the potential $ {\cal V}_{\cal R} $ is negative
(attractive) the perturbations are suppressed, while when it is
positive (repulsive) the perturbations are enhanced \cite{boy06}.
In the purely classical fast-roll case, $ {\cal V}_{\cal R} $ is
attractive and the perturbations are suppressed \cite{boy06}.

\medskip

In the slow-roll regime with a classical inflaton, the perturbation
modes eq.(\ref{ecmod}) feel the repulsive potential
$ W^{sr}_{\cal R} $ eq.(\ref{potW}) which yields the
scale invariant primordial power \cite{revmuk}
\be\label{ppBD}
P^{BD}_\mathcal{R}(k) = |{\Delta}_{k_0\;ad}^{\mathcal{R}}|^2 \;
\left(\frac{k}{k_0}\right)^{n_s - 1} \; ,
\ee
when Bunch-Davies (BD) initial conditions are imposed on the
perturbative modes.

In the classical fast-roll case the primordial power spectrum gets modified
by the transfer function  $ D_\mathcal{R}(k) $ as \cite{boy06,des08}
\be \label{ppmoD}
P_\mathcal{R}(k) = P^{BD}_\mathcal{R}(k)\left[1 +  D_\mathcal{R}(k) \right]
\ee
The effect of the precondensate regime on the primordial power
both for quantum slow-roll and quantum fast-roll initial conditions
can be also encoded in an appropriate transfer function
$ D_\mathcal{R}(k) $.

We will consider the effects on the primordial power at the end of
the transitory epoch before the effective classical slow-roll
epoch. The effects on the low CMB multipoles have two different
origins that come from the two kinds of transitories before
slow-roll:
\begin{itemize}
\item (a) {\bf Precondensate} effects. For perturbative $k$-modes that
exit the horizon {\bf before} the condensate eq.(\ref{efffield}) is formed
and before the classical equations (\ref{eqclasphi})-(\ref{epsilonclas})
hold, the transfer function $ D_\mathcal{R}(k) $ is nonzero and the
power spectrum is given by eq.(\ref{ppmoD}).
This entails an enhancement in the
low CMB multipoles as we show in sec. \ref{sec:qic}.
\item (b) {\bf Fast-roll} effects. As for the purely classical fast-roll
inflation initial conditions \cite{boy06,des08}, when the nonperturbative
$k$-modes obey quantum fast-roll initial conditions, a nonzero transfer
function $ D_\mathcal{R}(k) $ appears in eq.(\ref{ppmoD}) and the
primordial power is suppressed as we show in sec. \ref{sec:qic}.
\end{itemize}

The transfer function in the primordial power due to the deviation
from the slow-roll dynamics is given by
\be
D(k) = \frac1{k} \int_{-\infty}^0 d\eta {\cal V}_{\cal R}(\eta) \left[
\sin(2 \, k \; \eta) \left( 1- \frac1{k^2 \; \eta^2} \right) +
\frac2{k \; \eta} \cos(2 \, k \; \eta) \right] \; .
\ee
This function allows to obtain the changes (suppression or enhancement) in
the low multipoles resulting from the early fast-roll dynamics \cite{boy06}
\be
\frac{\Delta C_l}{C_l} = \frac{\int_0^\infty D(\kappa x)  \; f_l(x) \;
  dx}{\int_0^\infty f_l(x) \; dx}
\ee
where $ x=k/\kappa $, $ \kappa \equiv a_0 \; H_0 /3.3 $, with
$ a_0 $ and $ H_0 $ the scale factor and the Hubble parameter today,
and
\be
f_l(x) = [j_l(x)]^2 / x
\ee
for nearly scale invariant spectrum of perturbations \cite{boy06} with
$j_l(x)$ the spherical Bessel functions \cite{abramo}. In particular, for the
quadrupole this yields
\be
\frac{\Delta C_2}{C_2} = \frac1{\kappa}
\int_{-\infty}^0 d\eta \;  {\cal V}_{\cal R}(\eta) \;  \Psi(\kappa\eta) \; ,
\ee
with
$$
\Psi(x) \equiv 12 \, \int_0^{\infty} \frac{dy}{y^4} \; [j_2(y)]^2  \;
\left[ \left(y^2 - \frac{1}{x^2} \right) \sin(2 \, y \; x) + \frac{2y}{x}
\cos(2 \, y \; x) \right] \; .
$$
Analogously, tensor perturbation modes $ S_T $ verify the evolution equation
\be
\left[ \frac{d^2}{d\eta^2} + k^2 - W_T(\eta) \right]
S_{T}(k;\eta) = 0 \;, \quad \quad W_{T} = \frac{a''(\eta)}{a(\eta)} \; ,
\ee
with $ a(\eta) $ the scale factor as a function of the conformal
time and $ a''(\eta) $ its second derivative with respect to $ \eta $.

$ W_T(\eta) $ is the potential felt by the tensor perturbations
during the whole evolution: fast-roll as well as slow-roll inflation.

It is convenient to define the potential
\be
{\cal V}_T = W_{T} - W^{sr}_T \quad , \quad
W^{sr}_T \simeq \frac{2+3 \; \eta_v}{\eta^2} \; ,
\ee
where $  W^{sr}_{T} $ is the value of $ W_T $ in the slow-roll
evolution. The potential $ {\cal V}_T $ is
determined by the fast-roll dynamics.
When the potential $ {\cal V}_T $ is negative (attractive) the
tensor perturbations are suppressed, while when it is positive (repulsive)
the tensor perturbations are enhanced \cite{boy06}.
In the purely classical fast-roll case, $ {\cal V}_T $ is attractive
and the tensor perturbations are suppressed \cite{boy06}.

\section{Quantum initial conditions and observable consequences}
\label{sec:qic}

Several initial conditions have been shown to lead to inflationary
periods that last long enough ($ 55 $ efolds or more) to explain
the observed flatness and homogeneity of the Universe. Namely,
classical initial conditions that verify the classical
slow-roll conditions $ \dot\varphi^2 \ll m^2 \varphi^2 $,
quantum  initial conditions
that verify the quantum generalized slow-roll conditions
eq.~\eqref{gsrc}, and more recently classical fast-roll initial
conditions $ \dot\varphi^2 \sim m^2 \varphi^2 $
have been shown to lead to long enough inflation
\cite{cao04,boy06}. We show in this section that more general quantum
initial conditions lead to inflation. These initial
conditions are different from the quantum slow-roll conditions
and we call them quantum fast-roll initial conditions.
The classical fast-roll initial conditions are a particular case
of them.

\medskip

The observable consequences of slow-roll inflation are well known
\cite{infbooks} in particular those for the CMB fluctuations.
Present observational data agree with the slow-roll predictions
for appropriate inflaton potentials within the present accuracy
except for the quadrupole of the CMB. All initial conditions
mentioned here lead after a transitory epoch to a period of
slow-roll inflation. If the slow-roll inflation period is large
enough, all the details of the previous initial conditions will be
erased. However, if the larger scales exited the horizon close to
the end of the transitory epoch, the initial conditions imprint
features in the low multipoles of the CMB spectrum with respect to
the slow-roll results. Such corrections have been computed for the
case of classical fast-roll initial conditions in refs.
\cite{boy06,des08}. We compute here the corrections for the
quantum slow-roll conditions, and for the quantum fast-roll
initial conditions introduced in this paper. This computation
shows that the two possible types of transitory epochs present
before slow-roll, namely, fast-roll and precondensate, lead to
different kinds of corrections for the low CMB multipoles.

\subsection{Quantum slow-roll initial conditions}

\begin{figure}
\begin{center}
\includegraphics[scale=1.0]{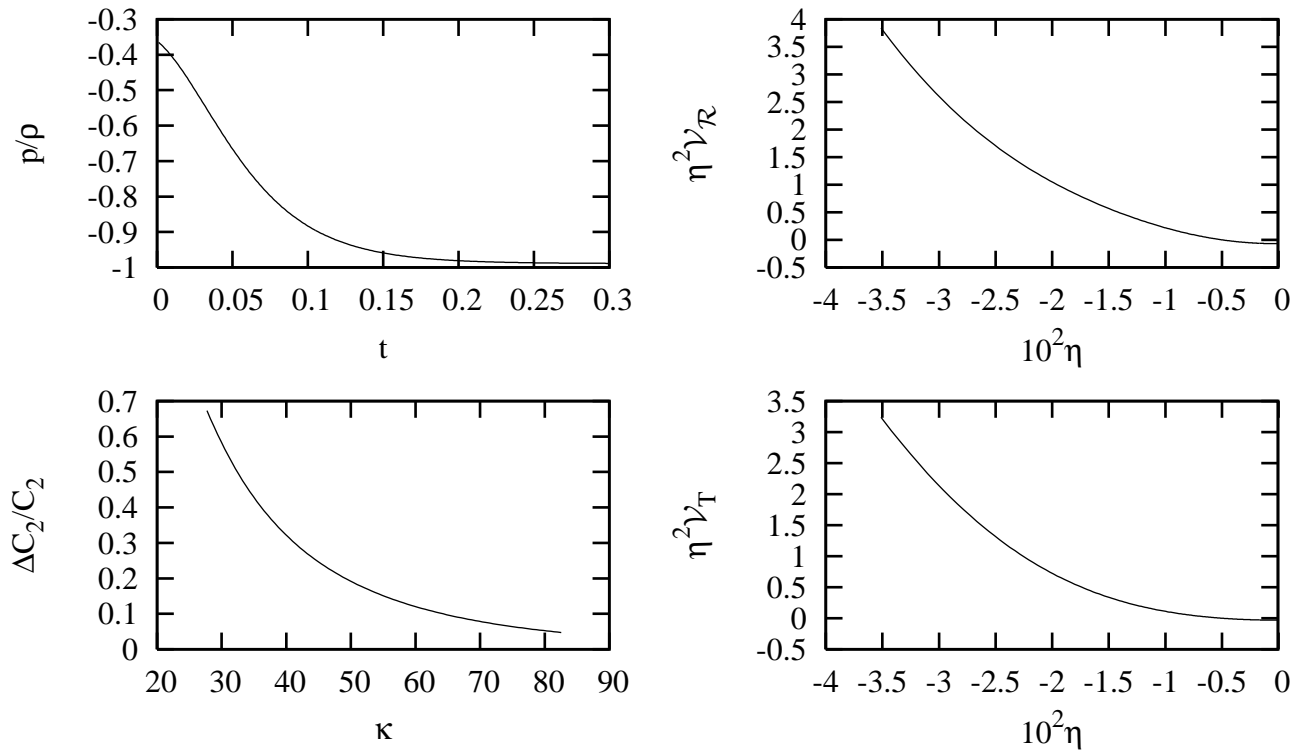}
\end{center}
\caption{Quantum chaotic inflation with \emph{quantum slow
roll initial conditions}. The nonperturbative inflaton modes have
wavenumbers around $ k \sim 8 \; m $ with a
characteristic width $ \Delta k \sim 0.8 \; m $.
Top left: equation of  state $ p/\rho $ as a function of comoving time for
early times. Top right: the potential $ {\cal V}_{\cal R} $ felt by the
scalar curvature perturbations as a function of the conformal time $ \eta $.
Bottom right: the potential $ {\cal V}_{T} $ felt by the tensor perturbations as a
function of the conformal time $ \eta $. Bottom left: the CMB quadrupole
amplitude, $ \Delta C_2 / C_2 $ as a function of the scale $ \kappa $.
$ \Delta C_2 / C_2 $ {\bf increases} due to the effect of the
quantum precondensate inflaton epoch. \label{tausrq8}}
\end{figure}

We first consider quantum initial conditions that verify the
quantum slow-roll condition eq.~\eqref{gsrc}. The background
dynamics of these quantum initial conditions has been studied in
ref.~\cite{tsuinf}. These initial conditions have initial
equations of state between $ p(0)/\rho(0) \simeq -1 $ [for
classical slow-roll, eq.~\eqref{csrc}] and $ p(0)/\rho(0) \simeq -
1/3 $ for the quantum extreme case. The latter equation of state
is found when the integral of eq.(\ref{mome}) dominates the energy
$ \rho $ [eq.~\eqref{H2rho}]. (See Fig.~\ref{tausrq8}).

\medskip

During the transitory precondensate inflationary
epoch the modes are rapidly redshifted and at the end of this epoch they are
assembled in a zero mode condensate. After that, the condensate epoch starts
and the quantum field condensate can be described by the effective
classical field eq.~(\ref{efffield}) that verifies the classical slow-roll
conditions. This allows
to recover the known slow-roll results for the cosmological relevant
modes provided they exited the horizon during the condensate epoch
\cite{cao04}. In addition, quantum corrections have been predicted in
Ref.~\cite{cao04} for large scales that exited the horizon
close to the transition between the quantum precondensate and the
classical condensate inflationary epoch.

Here, we numerically compute these quantum corrections to the
scalar and tensor metric perturbations. First, we compute the
potentials $ {\cal V}_{\cal R} $ and $ {\cal V}_{T} $ felt by
the scalar and tensor fluctuations, respectively.

\medskip

For quantum chaotic inflation we consider
$$ \frac{m}{M_{Pl}} = 1.02 \times 10^{-5} \sim
\left(\frac{M}{M_{Pl}}\right)^2
\quad {\rm and} \quad
\tilde\lambda = 1.18 \times 10^{-12} \sim \left(\frac{M}{M_{Pl}}\right)^4
$$
(best fit to WMAP3 and SDSS data found in Ref.~\cite{des08a}).

We consider quantum initial states that verify the
quantum slow-roll condition [eq.~\eqref{gsrc}] with various
wavenumbers distributions and with amplitudes chosen to lead to an
energy density at the beginning of slow-roll $ \rho
\simeq 6.5 \; m^4 / \tilde{\lambda} $. (These values guarantee that
the slow-roll part of inflation last $ 55 $ efolds). In
particular, in Fig.~\ref{tausrq8} we show the results obtained for
a distribution of nonperturbative inflaton modes around
$ k = k_0 \sim 8 \; m $ with width
$ \Delta k \sim 0.8 \; m $ and the slow-roll initial conditions
$$
\dot f_k(0)\simeq-\frac{k^2/a^2(0)+{\cal M}^2(0)}{3 \; H(0)} \; f_k(0) \; ,
$$
which follow by neglecting $ {\ddot f}_k(0) $ in eq.(\ref{eqNmodes})
\cite{tsuinf}.

\medskip

The results for quantum chaotic inflation show that {\bf both} the scalar and
the tensor metric perturbations became {\bf enhanced} by the quantum
slow-roll initial conditions at the large
scales that correspond to the end of the precondensate epoch.
Fig. \ref{tausrq8} illustrates this effect. In this case the potentials $ {\cal V}_{\cal
R} $ and $ {\cal V}_{T} $ felt by the fluctuations
are both {\bf repulsive}. This also implies that if
the scale corresponding to the CMB quadrupole exited the
horizon at the end of the precondensate epoch, its
amplitude will be larger than if it would exited deep inside the condensate
epoch. This enhancement is larger for initial
distributions whose energy is dominated by modes with higher
momenta. Since the CMB observations show a {\bf suppression} of the
quadrupole instead of enhancement, this simply implies that the CMB quadrupole mode
{\bf cannot} exit the horizon during a quantum slow-roll precondensate epoch. We will
see below that this statement gets weaker if the scale of the
quadrupole exited the horizon during a quantum fast-roll epoch.

\medskip

For quantum new inflation, on the other hand, initial conditions
have small amplitudes for the inflaton modes and for the expectation value of the
inflaton. This implies that even the more excited inflaton modes have small amplitudes, and
if they  have $ k \sim m $ the contribution of the integral eq.(\ref{mome})
to the energy density will be small, and also the precondensate corrections to the
spectrum will be small. In addition, if the initial conditions verify the quantum slow
roll condition, then the kinetic contribution
$$
\frac12 \, \dot\varphi^2 + \frac14 \int_R \frac{d^3k}{(2\pi)^3} \; |\dot f_k|^2
$$
is also small. This
implies that for new inflation if the more excited inflaton modes have $ k \sim m $ the
corrections to the scalar and tensor perturbations from the slow-roll precondensate
are negligible.

\subsection{Quantum fast-roll initial conditions}

We call quantum fast-roll initial conditions all those quantum initial
conditions that are different from the quantum slow-roll condition
eq.~\eqref{gsrc}. The classical fast-roll initial conditions
$ \dot\varphi^2 \sim m^2 \varphi^2 $ are particular cases of these
quantum fast-roll initial conditions.
We have found that quantum fast-roll
initial conditions can lead to inflation that last long enough, even when
their initial kinetic energy is of the order of their initial potential
energy. Generalizing the dimensionless variable $ y $ defined in
ref.~\cite{boy06} we define here the variable [see eq.(\ref{pressure})]
\be\label{defy}
y^2 \equiv \frac32\;\frac{p+\rho}{\rho}, \quad 0 \leq y^2 \leq 3.
\ee
Since inflation requires
\be
\frac{\ddot a}{a} = H^2 (1-y^2) > 0,
\ee
the range of the variable $ y^2 $ for inflationary evolution is
$ 0 < y^2 < 1 $.

\medskip

The classical slow roll regime corresponds to $ y^2 \ll 1 $. On the other
hand, $ y^2 \simeq 3$ can only be reached if the kinetic energy dominates
the energy density, in this case the equation of state is $ p \simeq \rho $.
We consider here quantum fast-roll initial conditions, that have kinetic
energy of the order of their potential energy and therefore $ y^2 \simeq 1 $.
In this case the equation of state is $ p \simeq -\rho/3 $.

\medskip

For quantum chaotic inflation we consider $ m/M_{Pl} = 1.02
\times 10^{-5} \sim \left(\frac{M}{M_{Pl}}\right)^2 $ and
$ \tilde\lambda = 1.18 \times 10^{-12} \sim \left(\frac{M}{M_{Pl}}\right)^4 $
(best fit to WMAP3 and SDSS data found in Ref.~\cite{des08a}).
We consider initial states obeying the fast-roll condition,
\be \label{dfnosr}
\dot f_k(0) \sim \sqrt{\frac{k^2}{a^2(0)} + {\cal M}^2(0)} \; f_k(0)
\ee
with various distributions and with amplitudes chosen to
lead to an initial energy density at the beginning of slow-roll
$ \rho \simeq 6.5 \; m^4 / \tilde{\lambda} $. (These values guarantee
that inflation lasts at least $ 55 $ efolds).
The conditions (\ref{dfnosr}) do not fulfil the quantum slow-roll conditions
eq.~\eqref{gsrc}.

\medskip

We find that quantum fast-roll initial conditions lead to a early
fast-roll epoch, in which the contribution from the
kinetic energy dominates the total energy density. During
this fast-roll epoch the large value of the Hubble constant overdamps the
derivatives of the inflaton modes and of the inflaton expectation value
rapidly reducing the kinetic energy. This makes that generically the
fast-roll epoch is followed by the slow-roll epoch.
In fact, it is in this slow-roll epoch where most of the required
efolds take place.

\begin{figure}
\begin{center}
\includegraphics[scale=1.0]{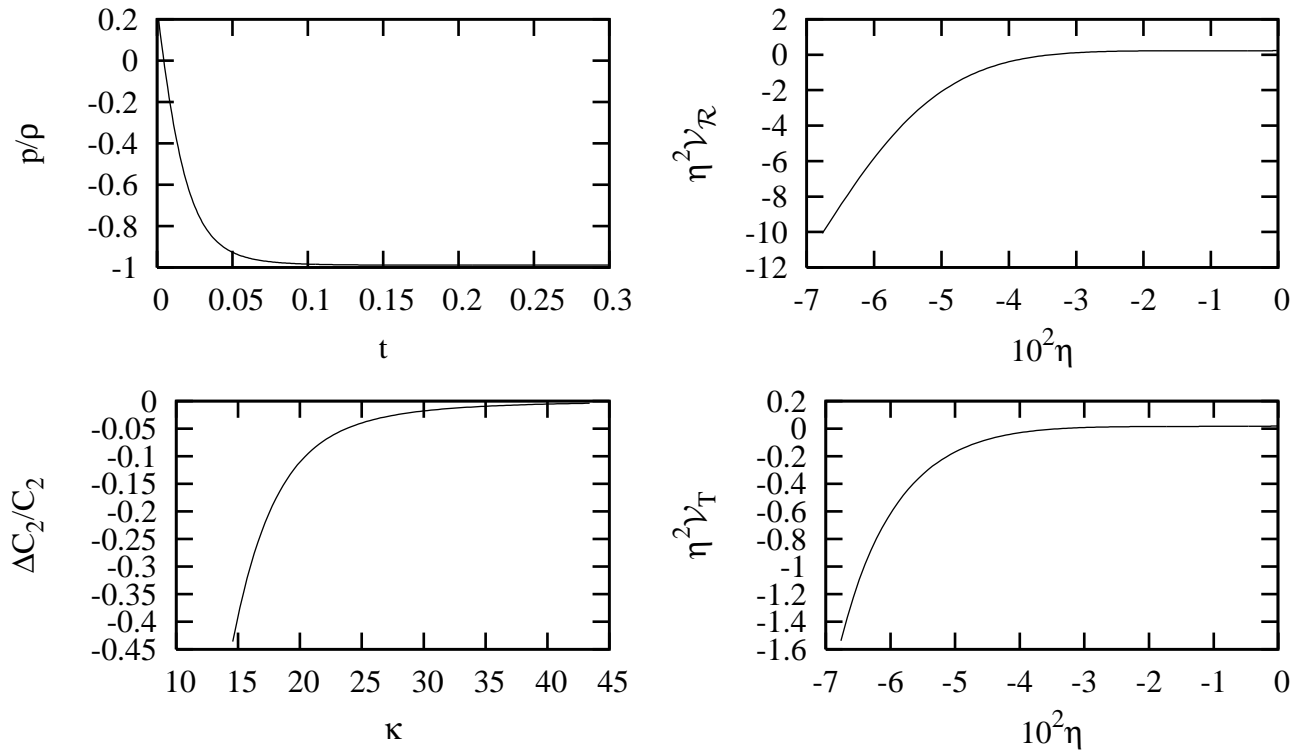}
\end{center}
\caption{Quantum chaotic inflation with \emph{quantum fast
    roll initial conditions} given by eq.~\eqref{dfnosr} with  low $ k $
inflaton excited modes having $ k \sim k_0 \ll m $. In this case the quantum
fast-roll stage is an effectively classical condensate epoch.
Top left: equation of state $ p/\rho $ as a function of comoving time for
early times.
Top right: The potential $ {\cal V}_{\cal R} $ felt by scalar curvature
perturbations
as a function of the conformal time $ \eta $. Bottom right: The potential
$ {\cal V}_{T} $ felt by tensor perturbations as a function of the conformal time $ \eta $.
Bottom left: change of the CMB quadrupole amplitude:
$ \Delta C_2 / C_2 $ as a function of the scale $ \kappa $.
The quadrupole gets suppressed
by the fast-roll condensate. \label{taunosrq0}}
\end{figure}

\begin{figure}
\begin{center}
\includegraphics[scale=1.0]{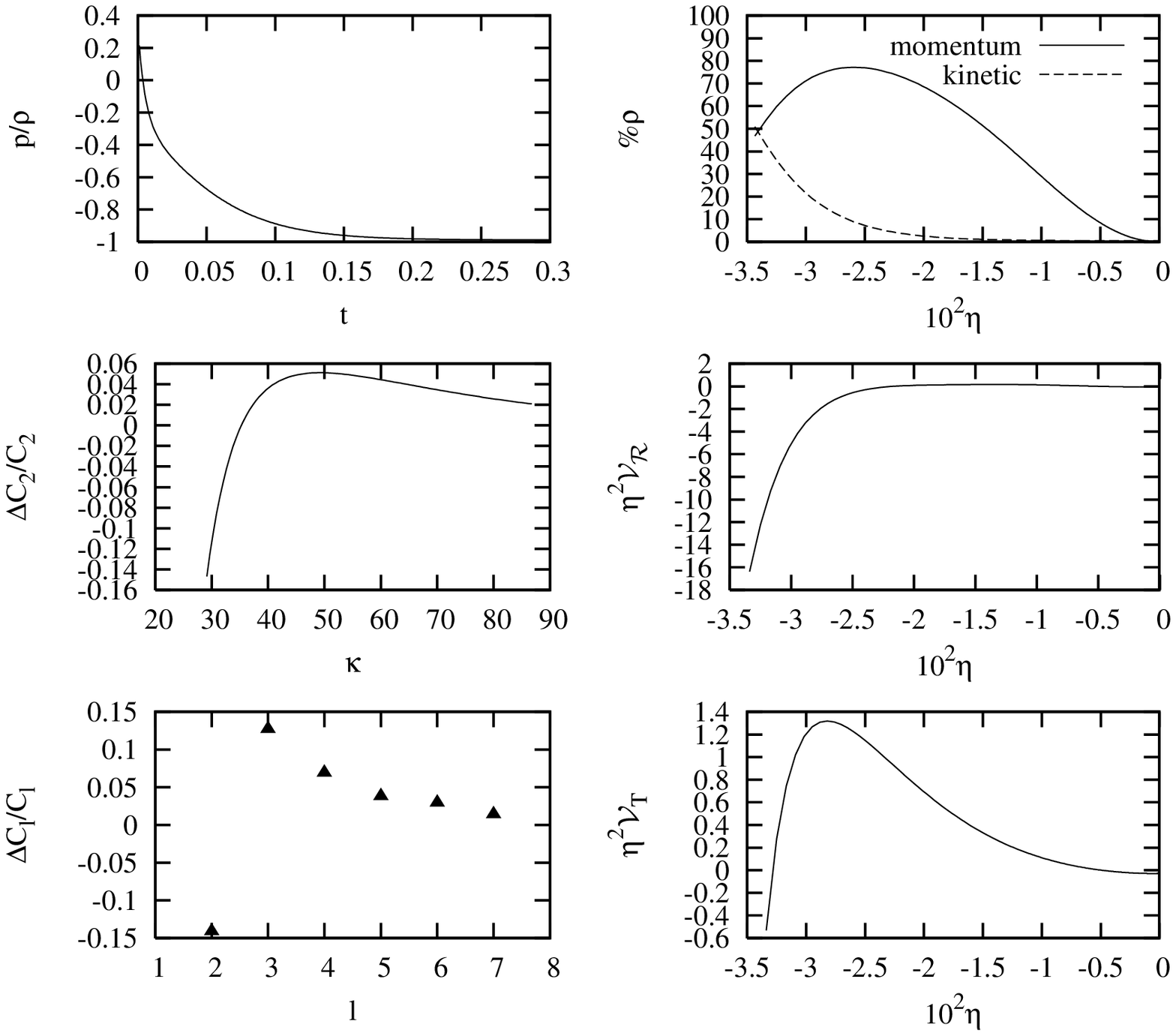}
\end{center}
\caption{Quantum chaotic inflation with \emph{quantum fast roll
initial conditions} given by eq.~\eqref{dfnosr} for excited
inflaton modes having $ k \sim k_0 \simeq 8 \; m $ and a
characteristic width $ \Delta k \sim 0.8 \; m $. Top left:
Equation of state $ p/\rho $ as a function of comoving time for
early times. Top right: Percentage ($\%$) fraction of energy $
\rho $ that comes from the momentum contribution [the integral
eq.(\ref{mome})] and from the kinetic term contribution (the
derivative of the mode amplitudes $ {\dot f}_k $) Middle right:
The effective potential felt by the scalar perturbations $ {\cal
V}_{\cal R} $ as a function of the conformal time $ \eta $. Bottom
right: The effective potential felt by the tensor perturbations $
{\cal V}_{T} $ as a function of the conformal time $ \eta $.
Middle left: Change of the quadrupole amplitude of the CMB, $
\Delta C_2 / C_2 $, as a function of the scale $ \kappa $. $
\Delta C_2 / C_2 $ gets suppressed or enhanced depending on the
value of the scale $ \kappa $. Bottom left: change of the low
multipoles of the CMB, $ \Delta C_l/C_l $, as a function of the
multipole order $ l $ for $ \kappa = 29.12 m $.\label{taunosrq8}}
\end{figure}

Two situations arise depending whether the initial distribution of
nonperturbative inflaton modes is centered around small or large $ k_0 $
($ k_0 \ll m $ or $ k_0 \sim m $):
\begin{itemize}
\item{ When the initially excited nonperturbative inflaton modes are
excited only for low wavenumbers $ k \sim k_0 \ll m $, they effectively
behave as a zero mode condensate, and we have an effective classical
description in terms of the effective inflaton field defined in
eq.~\eqref{efffield} with the evolution given by
eqs.~\eqref{eqclasphi}-\eqref{epsilonclas}. For these initial conditions
we recover the classical fast-roll epoch described in
ref.~\cite{boy06,des08}, followed by the classical slow-roll epoch,
in particular we have from eqs.(\ref{pressure}), (\ref{pclas}) and
(\ref{defy})
$$
y^2 =  \frac32 \; \frac{\dot{\tilde\varphi}_{eff}^2}{\tilde\rho} \; .
$$
Consistently, the corrections for the scalar and tensor metric
perturbations are the same as those found in Ref.~\cite{boy06,des08}.
Both scalar and tensor
metric perturbations that exited the horizon by the end of this effective
classical condensate fast-roll
epoch have their amplitudes {\bf suppressed} with respect to their values
for pure slow-roll implying a suppression of the low CMB multipoles,
in particular of the CMB quadrupole. See Fig.~\ref{taunosrq0}.}
\item{When the initially excited inflaton modes have $ k \sim k_0 \sim m $
or larger, the evolution starts with a quantum precondensate epoch, and
the fast-roll and quantum precondensate effects compete. Perturbation
modes that exit the horizon during
fast-roll are suppresed while exiting the horizon during
the precondensate enhances them. Their respective intensity
depends on the values of $ y^2 $ and of the characteristic $ k $
scale of the excited inflaton modes. This competition results in that at
some scales the perturbations are enhanced while at other scales
they are suppressed, as can be seen for example in
Fig.~\ref{taunosrq8}. The relative relevance between the kinetic
$ |\dot f_k|^2 $ and momentum contributions $ \frac{k^2}{a^2} \; |f_k|^2 $
to the energy [eq.(\ref{H2rho})] seems to determine which is the more
relevant effect at this $k$-scale, and whether the perturbations exiting
the horizon at this scale  $k$ are enhanced or suppressed relative to their
classical slow roll amplitude (see right panels of
Fig.~\ref{taunosrq8}). The enhancement is more notorious for the
tensor perturbations. In addition, the fact that the perturbations
at {\it some scales} are {\it enhanced} while at other  scales are
{\it suppressed}
implies that some CMB multipoles can be suppressed while others
are enhanced (see bottom right panel of Fig.~\ref{taunosrq8}).}
\end{itemize}

If the quantum initial precondensate effects would dominate when the
quadrupole mode exits the horizon, the quadrupole mode would get enhanced,
contrary to the CMB observations which show a suppression for the
quadrupole. Therefore, this imposes upper bounds on the characteristic
$ k_0 $ scale of the nonperturbative inflaton modes.
However, increasing the value of  $ y^2 $ reduces the enhancement of these
inflaton $k$-modes relaxing such upper bound on $ k_0 $.

\medskip

On the other hand, for quantum new inflation, the quantum initial
conditions require initially small amplitudes for the inflaton modes,
and this makes the contribution of the integral eq.(\ref{mome})
small, and therefore also the precondensate corrections to the spectra
will be small.

\medskip

We can summarize the time evolution for the quantum fast-roll inflaton
initial conditions as
\begin{itemize}
\item{(i) Early precondensate fast-roll epoch where the inflationary
dynamics is described by the nonperturbative inflaton $k$-modes
$ f_k(t) $.}
\item{(ii) The first of the two transitions happens. Either the fast-roll is
damped ending the fast-roll epoch, or the condensate is formed
ending the precondensate epoch. The initial conditions determine
what happens first. If the fast-roll is stopped first this second
epoch is a precondensate slow-roll epoch. While if the condensate
forms first this second epoch is a fast-roll condensate epoch
where the inflationary dynamics is described by the effective
inflaton $ \tilde{\varphi}_{eff}(t) $ of eq.(\ref{efffield}).}
\item{(iii) The second of the two transitions happens. The
dynamics reaches a slow-roll condensate regime from its previous
precondensate slow-roll epoch or fast-roll condensate epoch. This
slow-roll condensate evolves according to the effective
description given by eqs.(\ref{eqclasphi})-(\ref{epsilonclas}),
i.e., this third epoch is effectively a classical slow-roll
epoch.}
\end{itemize}

\section{Conclusions}

We have introduced a new type of inflationary initial conditions
namely quantum fast-roll initial conditions which lead to long
enough inflation. They are different from the quantum generalized
slow-roll conditions and yield to the classical fast-roll
conditions as a particular case. Quantum fast-roll initial
conditions are far beyond slow-roll as their kinetic energy can be
of the same order as their potential energy.

\medskip

Two relevant issues play an important role in the quantum inflaton
dynamics:
the redshift due to the expansion and the large value of the
Hubble parameter. The redshift due to the expansion succeeds to
suppress the contributions from the quantum inflaton $k$-terms, and
assembles all the nonperturbative inflaton modes in a inflaton condensate
after a transitory precondensate epoch. On the other hand, the large value
of the Hubble parameter succeeds to overdamp the oscillations of the
redshifted inflaton quantum modes thus stopping the fast-roll. Therefore,
the combined effect of the redshift due to the expansion and the large
values of the Hubble parameter makes that after a transitory quantum
inflation epoch an effective classical slow-roll inflation epoch is
reached.

\medskip

Such transitory epoch can have observable effects. If scales which
are today horizon-size exited the horizon close to the end of the
transitory epoch, the spectra of scalar and tensor metric
perturbations are similar to the spectra for classical slow-roll
but with corrections for the larger scales. These corrections for
the larger cosmologically relevant scales of the scalar and tensor
metric perturbations are reflected in the amplitudes of the lower
multipoles of the CMB spectra. We have found that for both scalar
and tensor metric perturbations, the classical or effective
classical condensate fast-roll epoch {\bf suppresses} the
amplitude of the perturbations (and therefore the amplitude of the
respective CMB multipoles), while the quantum precondensate epoch
{\bf enhances} the amplitude of the perturbations (and of the
respective CMB multipoles).

The corrections to the CMB multipoles from the quantum precondensate
turn to be smaller in new inflation than in chaotic inflation.

In addition, when both fast-roll and quantum precondensate effects compete
they can lead to suppression in some scales and enhancement in
other scales. Therefore, these corrections for the lower
multipoles can allow to further improve the fit of the predictions
of the inflation plus the $\Lambda$CMB model to the observed TT, TE and EE
CMB spectra.

\medskip

It is important to note that these corrections to the spectrum of
scalar metric perturbations arise as natural consequences
of the quantum dynamics of the inflaton. Since the quantum dynamics also
predicts
corrections to the spectrum of tensor perturbations, when the
tensor perturbations spectrum will be observed, an independent test
of the quantum fast-roll model will be possible.
Therefore, effects due to quantum fast-roll initial conditions
in the nonperturbative inflaton modes ($ k > \Lambda $)
are a consistent and contrastable
explanation for the suppression of the CMB quadrupole and for the other
departures of the low CMB multipoles from the slow-roll inflation plus
$\Lambda$CMB model.

The consequences of our study of quantum fast-roll initial conditions
on the low TE and EE multipoles need to be explored. Forthcoming data
for low CMB multipoles motivate this work and
are needed to reach a clear understanding on these
issues.

\acknowledgments

F.J.C. thanks the LPTHE (Paris) and the Observatoire de Paris
for their hospitality, and
acknowledges support through Research Projects Nos. FIS2006-05895
(MECD, Spain), ESP2007-30785-E (MECD, Spain), and CCG07-UCM/ESP-2925
(UCM/CAM, Spain).

\end{document}